%% file: SnowmassBook.tex
\documentclass{tcibook}
\usepackage{fancyhea}
\usepackage{work}
\usepackage{bm}       
\usepackage{graphicx}
\usepackage{multirow}
\usepackage{lineno}
\usepackage{threeparttable}
\usepackage[normalem]{ulem}
\usepackage{color}
\usepackage{xspace}
\usepackage{siunitx}
\usepackage[colorlinks = true,
            linkcolor = blue,
            urlcolor  = blue,
            citecolor = blue,
            anchorcolor = blue,
            hypertexnames=false] 
            {hyperref}
\usepackage{amsmath}
\usepackage{amssymb}
\input workshopsymbols.tex     
 
\newcommand{\ndbd}{$0\mathrm{\nu\beta\beta}$\xspace}

\newcommand{\hpge}{HPGe}
\newcommand{\icpms}{ICP-MS}

\def\authorlist#1#2{
    \vskip 0.4in
\begin{center}\begin{large} {\bf  #1 } \end{large}
    \vskip 0.2in
              #2
     \vskip 0.2in
   \end{center}
}

\begin{document}


\pagenumbering{roman}

\parindent=0pt
\parskip=8pt
\setlength{\evensidemargin}{0pt}
\setlength{\oddsidemargin}{0pt}
\setlength{\marginparsep}{0.0in}
\setlength{\marginparwidth}{0.0in}
\marginparpush=0pt



\renewcommand{\chapname}{chap:intro_}
\renewcommand{\chapterdir}{.}
\renewcommand{\arraystretch}{1.25}
\addtolength{\arraycolsep}{-3pt}

\thispagestyle{empty}
\begin{centering}
\mbox{\null}
\rightline{\begin{tabular}{l}
FERMILAB-CONF-xx\\
SLAC-PUB-xx\\
 \end{tabular}}
\vfill

{\Huge\bf The Future of US Particle Physics}

\vskip 0.6in

{\LARGE  Report of the 2021  US  Community Study  \\
     on the Future of Particle Physics

                  \smallskip

       organized by the  APS Division of Particles and Fields
       
                   \bigskip
                   \bigskip
                   \bigskip
                   \bigskip
                   
    \bf Topical Report:  Underground Facilities for Neutrinos}

\vfill

\vfill

\end{centering}




\pagenumbering{roman}






\tableofcontents

 \pagenumbering{arabic}

\input Underground/UF01/Neutrinos.tex





\end{document}

%% file: workshopsymbols.tex
\def\beq{\begin{equation}}
\def\eeq#1{\label{#1}\end{equation}}
\def\eeqn{\end{equation}}

\newenvironment{Eqnarray}%
   {\arraycolsep 0.14em\begin{eqnarray}}{\end{eqnarray}}
\def\beqa{\begin{Eqnarray}}
\def\eeqa#1{\label{#1}\end{Eqnarray}}
\def\eeqan{\end{Eqnarray}}

\let\bar=\overbar

\def\lsim{\mathrel{\raise.3ex\hbox{$<$\kern-.75em\lower1ex\hbox{$\sim$}}}}
\def\gsim{\mathrel{\raise.3ex\hbox{$>$\kern-.75em\lower1ex\hbox{$\sim$}}}}

\def\L{{\cal L}}

\def\W{{\cal W}}
\def\H{{\cal H}}

\def\del{\partial}
\def\Dslash{\not{\hbox{\kern-4pt $D$}}}
\def\dslash{\not{\hbox{\kern-2pt $\del$}}}
\def\pslash{\not{\hbox{\kern-2pt $p$}}}
\def\ETmiss{\not{\hbox{\kern-4pt $E$}}_T}

\def\Dlr{\mathrel{\raise1.5ex\hbox{$\leftrightarrow$\kern-1em\lower1.5ex\hbox{$D$}}}}

\def\MSB{{\bar{M \kern -2pt S}}}
\def\msb{{\bar{\scriptsize M \kern -1pt S}}}

\def\drb{{\bar{\scriptsize D \kern -1pt R}}}

\def\eV{{\rm eV}}
\def\keV{{\rm keV}}
\def\MeV{{\rm MeV}}
\def\GeV{{\rm GeV}}
\def\TeV{{\rm TeV}}

\def\s#1{\widetilde{#1}}



%% file: Underground/UF01/Neutrinos.tex
\setcounter{chapter}{0} 


\chapter{Underground Facilities for Neutrinos}
\label{ch:ufndbd}
\authorlist{T.~Bolton, M.P.~Decowski, A.~De~Roeck$^*$, G.~Orebi~Gann$^*$, D.~Speller \newline \textit{$^*$Liaisons from other Snowmass Frontiers.} }
   {(contributors from the community: Dongming Mei and others)}

\section{Accelerator neutrinos}
\label{sec:accelnu}
The long baseline neutrino physics plan for the US for the next two decades is expected to largely follow from the 2014 P5 process. The science case is summarized by the Neutrino Frontier, and this section will accordingly focus on logistics and timing. \ Data taking will continue through most of the decade for the No$\nu$a experiment, using neutrino beams directed from Fermilab to the Ash River, Minnesota far detector; and the T2K experiment utilizing the J-PARC neutrino beam on the east coast of Japan and the SuperK far detector in Kamiokande in the mountains of western Japan. \ In Japan, construction of HyperK has begun and will complete in the latter part of the decade. \ The US program is dominated by the Deep Underground Neutrino Experiment, DUNE. \ Currently the involvement of the US\ community in HyperK is limited, and future involvement is unclear at this time. \ 

\ The international DUNE Collaboration is currently building the components of
the first phase of the DUNE experiment: two 17 kiloton liquid argon TPC
modules comprising the far detector to be installed at the Sanford Underground
Research Facility (SURF) in Lead, South Dakota; and a multi-component near
detector to be installed at Fermilab. \ The US contributions to DUNE are
managed through the DOE's LBNF/DUNE-US project, which is responsible for
excavating and outfitting laboratory space at SURF, for building the required
conventional facilities at Fermilab, including a new wide band neutrino beam
that will initially operate at 1.2 MW\ of proton beam power, and for
contributing US components to the near and far detectors. \ 

LBNF/DUNE-US is the largest project ever undertaken by the DOE\ Office of High
Energy Physics, with an estimated cost of approximately 3.1 billion dollars.
\ International partners contribute half the resources required for the DUNE
detectors and provide substantial in-kind contributions to the facilities.
\ Notably and historically, CERN is providing both of the $66\times19\times18$
m$^{3}$ far detector liquid argon cryostats. Mega-science projects built in
the US have faced budget and schedule challenges, and LBNF/DUNE-US is no
exception. \ Nonetheless, all signs to date indicate strong support for DUNE
in the international neutrino science community, the DOE and partner
international science agencies, and the executive and legislative branches of
the US\ government. For example, in March 2022 the DOE Office of Science
committed to a funding profile for LBNF/DUNE-US that supports the project's
budget and keeps it on a competitive timeline. This discussion will assume
that such support continues, and that DUNE will come online at SURF in its
phase 1 configuration by 2030.

Perhaps the most interesting issue for DUNE in the Snowmass Underground
Frontier context is its upgrade timetable, given that its first phase physics
program is largely set. The LBNF/DUNE-US\ project will excavate space for four
17 kiloton liquid argon TPCs in two large ($>140$ m long) detector caverns,
along with a central utility cavern between the two. \ Two of the slots will
be left empty for the first phase of DUNE, leaving available considerable
valuable scientific real estate. \ The \ DUNE collaboration would like to
proceed to its full phase 2 configuration as soon as possible. \ This entails
outfitting the SURF caverns with the third and fourth far detector modules,
upgrading the near detector at Fermilab, and increasing the Fermilab proton
beam power to 2.4 MW. \ However, the timeline for phase 2 is unclear, and one
could consider use of the vacant DUNE cavern space for other science, either
in cooperation with or independently of DUNE. \ Focusing on the far site at
SURF, one could imagine several possible scenarios:

\begin{enumerate}
\item Funding would allow early\ construction of the two remaining DUNE far
detector modules in, e.g., the 2032-2040 time frame. \ Practically speaking,
given the limited time for further development, DUNE may have to stick close
to one of its two existing detector technologies, \textquotedblleft horizontal
drift\textquotedblright\ or \textquotedblleft vertical drift\textquotedblright%
\ single phase liquid argon TPCs. \ DUNE would benefit from the ability to
acquire larger data sets quicker, but it might  gain only modest new science
capabilities, for example, by pushing energy measurement thresholds low enough
to do solar neutrino physics. Furthermore, the ability of DUNE to utilize
larger data sets for long baseline oscillation physics will depend on the
ability to control systematic uncertainties with its near detector, which
could result in prioritizing near detector upgrades at Fermilab.

\item Perhaps more plausibly, the final two DUNE far detector modules would be
installed at some intermediate later date, e.g., 2036-2044. \ This would
permit more time for development of new detector capabilities that could
expand the DUNE science program. \ More intriguingly, the extra time could be
used to design multi-use modules, i.e., ones that could support the DUNE long
baseline physics while enabling physics beyond the original DUNE scope, such
as neutrino-less double beta decay or direct dark matter searches.

\item If funding for the final DUNE far detector modules takes longer to
appear, then new far detector module installation could be pushed beyond 2040.
\ If this were the case, there may be time to build, execute, and dismantle
experiments dedicated to neutrino-less double beta decay, direct dark matter
searches, or other physics that utilize the LBNF cavern space but otherwise
take place outside of the DUNE framework.
\end{enumerate}

Which, if any, of the scenarios described above actually plays out is anyone's
guess. \ Any path chosen would obviously require extensive consultation among
multiple stakeholders. \ Given the scale, cost, and complexity of the choices,
guidance from the DOE Office of Science will likely be essential. \ Indeed,
one could imagine a recommendation emerging from the ongoing Snowmass/P5
process along the lines of \textquotedblleft Establish a clear and transparent
process to optimize the scientific utilization of excavated underground spaces
at SURF.\textquotedblright

From the point of view of the Underground Frontier, the most important things
to identify are requirements for any detectors placed in the DUNE caverns
beyond the envisioned liquid argon TPC units. \ Several specific items should
be assessed early:

\begin{enumerate}
\item No extraordinary procedures have been adopted for radiological
mitigation by the LBNF/DUNE-US project. \ Hence, extra shielding required in
the caverns, requirements for low radioactivity cryostat materials, radon
mitigation systems, or other similar items would need to be planned for.

\item The two DUNE detector caverns are supported by a large central utility
cavern that has been designed for four 17 kiloton liquid argon TPCs. \ Any
additional utility requirements needed for possible detectors that goes beyond
what is provided for DUNE (cryogenics, power, cooling, etc.) need to be
identified. \ Utilities could include items such as clean rooms or clean spaces.

\item Installation of any new detectors in the DUNE caverns would occur during
operations of the first two DUNE modules, and conceivably other experiments in
the SURF complex. \ Required access to the two SURF shafts, underground
occupancy limits, and other working conditions would need to be understood.

\item Surface facilities at the SURF\ site are limited. \ These facilities
include laboratory and office space, as well as the essential technical,
business, and ESH services required to support an underground science program. \ 
\end{enumerate}

\section{Non-accelerator neutrinos}
\label{sec:astrondbd}
The accelerator-based program to investigate the properties of neutrinos with the DUNE experiment (Section \ref{sec:accelnu}) 
is a critical part of maintaining U.S. world leadership in the field.  However, DUNE is not optimized to provide sensitive tests of the Majorana nature of the neutrino, or of its absolute mass\textemdash two high priority measurements which, as a result, require alternative experimental techniques.  Additionally, our understanding of neutrino properties has a direct impact on the interpretation of data in astrophysics and geophysics, where the emission of neutrinos carries important information about the structure and makeup of inaccessible systems.  To address these needs, the accelerator neutrino program must be strongly complemented by sensitive tests for symmetry violations (most notably neutrinoless double-beta decay, \ndbd) and by dedicated studies of neutrinos from natural sources. 

While individual non-accelerator neutrino experiments have distinct physics goals, the infrastructure requirements often strongly overlap.  This includes the need for facilities capable of hosting and supporting large-exposure, low-background particle detectors underground.

\subsection{Neutrinos from natural sources}
The detection of neutrinos from geological and astrophysical sources is a unique opportunity to collect data from systems that are otherwise practically inaccessible, ranging from stellar interiors to the activity beneath the earth's mantle.  
Science drivers for these experiments and observatories typically focus on far-reaching neutrino sources, including neutrinos from cosmological, stellar, geological, atmospheric, and extragalactic origins, as well as from systems like core-collapse supernovae and the diffuse supernova background \cite{UF1-nf042022}.  In addition to providing access to these systems, neutrino observations also provide the opportunity to address fundamental questions about neutrino properties.  These observations provide constraints on neutrino behavior and interactions with other constituents of matter, the number of neutrinos, the sum of the neutrino masses, and data on collective oscillations.  In turn, these studies yield new insight into flavor physics, new interactions, and often provide enhanced sensitivity to non-standard physics. This field has undergone significant growth since the last Snowmass exercise, and is likely to continue to grow with the recent rise of multi-messenger astronomy \cite{U1-multimess2021}. 

\paragraph{Progress since 2013}
Since 2013, several major experiments and observatories for neutrinos from natural sources (e.g. neutrinos of cosmological, astrophysical, and geological origin) have begun data collection, produced results, or completed operations.  Some of the experiments (for example, CLEAN) have as their primary aim the detection of dark matter or other new physics.  Among the experiments with US participation that have been completed since the last report are LVD \cite{U1-agafonova2022}, Borexino\cite{U1-borexino08}, and SNO\cite{U1-sno2013}. Ongoing experiments include Super Kamiokande \cite{U1-gado0abe2022}, CLEAN \cite{U1-clean2005}, and KamLAND/Kamland-Zen \cite{U1-kamlandgeo, U1-kamlandzen800}; while several major next generation experiments are under development and construction, including Hyper-Kamiokande \cite{U1-hyperk-ptep2015}, JUNO \cite{U1-juno2016}, and SNO+\cite{U1-snoplus2021}. 

Longer-term proposed experiments include a multi-tonne scale liquid xenon experiment for the detection of natural neutrinos \cite{U1-g3lxeloi,U1-xiangloi}, and a large hybrid scintillator detector called \textsc{Theia}~\cite{U1-theiawp}.  For a full discussion of the science goals of next-generation astrophysical and geophysical neutrino studies we refer the reader to the relevant frontier report \cite{UF1-nf042022}.

\paragraph{Honorable Mention: Underwater Neutrino Physics Experiments}
Several additional neutrino facilities exist worldwide, not just underground, but beneath water and ice. The ANTARES and KM3NeT Neutrino Telescopes are both in the Mediterranean (ANTARES ended data-taking as of February 2022)~\cite{U1-spiering2020}, while the US-led IceCube detector is in Antarctica \cite{U1-halzen2021}. A planned extension to IceCube known as IceCube-Gen2 is expected to begin taking first physics data after 2025 \cite{U1-spiering2020}. While these experiments play a major role in the future of US neutrino and high energy physics, requirements associated with their facilities lie beyond the scope of this report.

\subsection{Underground facilities for neutrinos from natural sources}

There are few experiments solely dedicated to studies of neutrinos from natural sources.  The variety of possible sources is wide, and interactions of interest occur over a broad range of energies.  At low neutrino energies (a few MeV and below), such as for terrestrial and cosmological neutrino sources, the needs of natural source experiments overlap with those of searches for \ndbd. Above a few MeV, experimental needs overlap with those of long-baseline neutrino experiments, while at higher energies, the needs overlap with those of dark matter experiments due to elastic scattering and coherent elastic neutrino-nucleus scattering (CEvNS).
A large fraction of natural source neutrino studies are conducted using large, multipurpose experiments that transcend categories, or using detectors designed to primarily  target other searches such as \ndbd.  Because of the strong overlap between the underground facilities needs of experiments for neutrinos from natural sources and those for neutrinoless double-beta decay at low energies, a discussion of the needs both, with some emphasis on the context of \ndbd, can be found in Section \ref{sec:uf4ndbd}. A few needs specific to astrophysical, geological, and reactor neutrinos are outlined below, while experiments responding to the supporting facilities survey are shown in Table \ref{tab:nuastrouf}.


\begin{table}[h]
    \centering
    \begin{tabular}{|p{6cm}|}
    \hline 
          Astrophysical/Natural \\
        \hline
          Hyper-Kamiokande  (Current)  \\
              Super-Kamiokande  (Current)  \\
              SNO+ (Current) \\
              KamLAND (Current) \\
              G3 Liquid Xenon Detector (LOI) \\
              \textsc{Theia} (LOI) \\
        \hline
    \end{tabular}
    \caption{Astrophysical and geological neutrino experiments responding to the UF4 underground supporting capabilities survey. Comments in parenthesis indicate whether the experiment is currently in progress or in development (LOI).}
    \label{tab:nuastrouf}
\end{table}

\subsubsection{Future needs and expectations for facilities}
For experiments involving large amounts of noble liquids, there are several challenges in moving to larger scales, including siting and space, depth, and cleanliness.  Aside from the primary detector, space for research and development, prototyping, and the operation of demonstrators is needed in order to properly characterize detector modules prior to installation.   

There is a need for new space for large detectors.  \textsc{Theia} is a large-scale natural source experiment proposed as a new DUNE module, or a standalone detector at the same site.  \textsc{Theia}'s science program is substantially enhanced by accessing the LBNF neutrino beam at Sanford laboratory.  In order to locate the experiment at SURF and to best take advantage of the beam and the deep site, a new, large cavern would need to be developed.

A kiloton scale detector (for example, a 35m $\textrm{GXe}_\textrm{nat}$ detector for neutrino studies and neutrinoless double-beta decay \cite{U1-ktxeprd2021})
    would also require new space.  An unenriched gaseous detector of xenon would be approximately as large as Super Kamiokande, with additional space required for shielding and utilities. These considerations must be taken into account in facilities planning.
    
For geophysical neutrinos, experiments are often constrained to sites where infrastructure exists.  More flexibility in underground locations would enhance global geoneutrino programs.

In order to acquire sufficient amounts of liquid noble gases for future large-scale experiments, R\&D paths are needed for the extraction, production, and storage of Xe and underground Ar.  Facilities support for the safe handling of large volumes of liquid cryogens underground will require careful planning.







    
    

\subsection{Neutrinoless double beta decay}
\label{NDBD}


Both the US and the international nuclear and particle physics community recognize neutrinoless double-beta decay (\ndbd) as one of the highest-priority, non-accelerator-based searches with significant need for underground detector facilities.  This priority has been repeated in meetings such as the North America-Europe Workshop on Double Beta Decay \cite{U1-naeudbd2021}, following which laboratory directors, funding agencies, and leaders across countries voiced support for the establishment of next generation \ndbd detectors on both continents.
The experimental investigation of the neutrino as a Majorana particle is an essential ingredient in the effort to understand the properties of neutrinos, and the search for neutrinoless double-beta decay remains one of the most sensitive probes of neutrino properties and the Majorana nature of the neutrino mass. 
Most searches involve the measurement of the half-life of a decay-capable isotope through detection of a small excess in the event rate at the energy (Q-value) of the decay. 
The recent improvement in sensitivity of \ndbd
experiments, and the successful completion of several of those experiments, has demonstrated both the progress of the field and the need to plan for the future requirements of large-scale $0\nu\beta\beta$ searches.  
 
The ``timely development and deployment of a US-led ton-scale neutrinoless double-beta decay experiment'' was also named as one of four high priority recommendations of the 2015 Long Range Plan for Nuclear Science~\cite{U1-2015lrp}.
The science drivers for \ndbd are discussed extensively in the Neutrino Properties Topical Group Report~\cite{U1-nf052022}, and can be briefly summarized as: 
\emph{
\begin{itemize}
    \item Is the neutrino a Majorana fermion or a Dirac fermion? 
    \item How can the sensitivity of neutrinoless double-beta decay searches best be improved beyond the inverted-ordering region targeted by the next generation of experiments?
\end{itemize}
}

Reaching the detector sensitivities required to definitively answer these questions requires, among other things, a clear strategy for background mitigation, which includes underground operation.  The community must begin planning now to accommodate the depth, space, and infrastructure requirements for experiments beyond the tonne-scale.

\paragraph{Progress Since 2013}


\label{sec:ndbd-recentupdates}
Since the last report in 2013, a number of mid-scale ({$\mathcal{O}(\SI{100}{\kilogram}$}) of isotope) experiments have been completed, placing increasingly stronger limits on 
\ndbd-capable isotopes.   
Interpreting these results in the neutrino-mixing framework, puts the limits of these mid-scale experiments at or near the top of the so-called Inverted Mass-Ordering of neutrinos, and also constrains the effective Majorana mass as a function of the mass of the lightest neutrino.  For further details, see the the NF05 report \cite{U1-nf052022}.

\begin{table}[hbtp]
    \centering
    \begin{tabular}{|c|c|c|c|c|}
    \hline
         Experiment & Isotope  & Location  \\
    \hline
        CUPID-0 \cite{U1-cupido2019} &  $^{82}\mathrm{Se}$  & LNGS, Italy\\
        CUPID-Mo \cite{U1-cupidmo2022} & $^{100}\mathrm{Mo}$  & Canfranc, France \\
        EXO-200 \cite{U1-exo200-2019} & $^{136}\mathrm{Xe}$  &  WIPP, USA\\
        \textsc{Gerda} \cite{U1-gerda} & $^{76}\mathrm{Ge}$ &    LNGS, Italy \\
        KamLAND-Zen 400 \cite{U1-kamlandzen400-2012} & $^{136}\mathrm{Xe}$  & Kamioka, Japan \\
        MAJORANA Demonstrator \cite{U1-mjd2022} & $^{76}\mathrm{Ge}$  & SURF, US \\
        \hline
    \end{tabular}
    \caption{Examples of of recently completed experiments with major US leadership or participation. 
    }
    \label{tab:usndbdexper}
\end{table}
%
\begin{table}[hbtp]
    \centering
    \begin{tabular}{|c|c|c|c|c|c|}
    \hline
        Experiment & Isotope &  Location \\
    \hline
        CUORE \cite{U1-cuore2022} & $^{130}\mathrm{Te}$   & LNGS, Italy\\
        AMORE-I \cite{U1-amorefam} & $^{100}\mathrm{Mo}$ & Yangyang, South Korea \\
        KamLAND-Zen 800 \cite{U1-kamlandzen800} & $^{136}\mathrm{Xe}$  & Kamioka, Japan \\
        LEGEND-200 \cite{U1-legend} & $^{76}\mathrm{Ge}$  & LNGS, Italy \\
       NEXT-xxx(white, 100, tonne-scale) & $^{136}\mathrm{Xe}$  & LSC, Spain \\
        SuperNEMO Demonstrator \cite{U1-supernemo2010, U1-supernemomethod2021} & $^{82}\textrm{Se}$ & LSM, France\\
        SNO+ \cite{SNO:2021xpa} & $^{130}\mathrm{Te}$  & SNOLAB, Canada\\
    \hline     
    \end{tabular}
    \caption{Ongoing experiments with US leadership or participation.  
    }
    \label{tab:usndbdrunning}
\end{table}

The experimental search for $0\nu\beta\beta$ is an international effort, with strong US participation in a number of completed (Table \ref{tab:usndbdexper}) and ongoing (Table \ref{tab:usndbdrunning}) searches.
The results of these searches compare competitively with other worldwide efforts such as the completed CANDLES-III\cite{U1-candles3} ($^{48}\mathrm{Ca}$ at Kamioka, Japan) and CDEX ($^{76}\mathrm{Ge}$, at CJPL, China), and the ongoing experiments 
 COBRA ($^{116}\mathrm{Cd}$ at LNGS, Italy),
 SuperNEMO\cite{U1-supernemomethod2021} ($^{82}\mathrm{Se}$ at LSM, France),
 PandaX-III\cite{U1-panda3} ($^{136}\mathrm{Xe}$ at CJPL, China), and 
 AMoRE /AMoRE II\cite{U1-amorefam} ($^{100}\mathrm{Mo}$) at Yangyang, South Korea).  In addition, next-generation \ndbd detectors have been proposed that build on cross-collaborative strengths by combining efforts, including \textsc{Legend}\cite{U1-legend} (\textsc{Gerda}\cite{U1-gerda} and \textsc{Majorana}\cite{U1-mjd2022}) and nEXO\cite{U1-nexo}.  

The experimental techniques break down in roughly four categories: 
\begin{itemize}
    \item Large-scale liquid scintillators doped with the \ndbd isotope, e.g. SNO+ and KamLAND-Zen experiments
    \item Gas and liquid time-projection-chambers filled with the \ndbd isotope, e.g. nEXO and NEXT experiments
    \item Semiconductors made from the \ndbd isotope, e.g. LEGEND
    \item Bolometers attached to crystals made from the \ndbd isotope, e.g. CUPID and CUORE  
\end{itemize}

LEGEND, nEXO, and CUPID \cite{U1-cupidprecdr} (the CUORE Upgrade with Particle Identification) are now three proposed experiments with major US involvement poised to begin either or both of commissioning and data collection within the next several years, and represent a concentration of US resources in extending the search for \ndbd to half-lives greater than $10^{28}$ yr.  The infrastructure and supporting capabilities necessary to undergird these efforts are key aspects of the planning process for each experiment and the laboratories that host them. 

\begin{table}[h]
    \centering
    \begin{tabular}{|p{6cm}|}
    \hline 
         \ndbd  \\
        \hline
             nEXO (Planned)   \\
             Majorana Demonstrator (Current)\\
             NEXT-100 (Planned)   \\            SNO+ (Current) \\
             A possible \ndbd extension to DUNE (Conceptual) \\
             NuDot (Current + Planned)             \\
             Kiloton Xe TPC for \ndbd  (Conceptual)       \\
             CANDLES (current + planned)                      \\
             NEXT-CRAB (Planned)                      \\      
             NEXT-HD (Planned)                        \\  
             NEXT with Ba-Tagging (Planned)            \\  
             KamLAND-Zen (Current)                    \\
             CUPID  (Planned)                           \\
             LEGEND (Planned)                     \\
             \textsc{Theia} (Planned)           \\
        \hline
    \end{tabular}
    \caption{Neutrinoless double-beta decay experiments responding to the underground survey}
    \label{tab:ndbduf}
\end{table}

\subsection{Underground Facilities for \ndbd 
}
\label{sec:uf4ndbd}


The depth requirement for \ndbd and natural source experiments arises from the need for shielding from cosmic ray backgrounds.  While some proposals for virtual depth enhancement have been proposed (e.g. through active background suppression \cite{U1-wiesinger2018}), such techniques are still under development, and suppression of backgrounds by physically siting deep underground is still critical for the science reach of these experiments.  Ideally, this consideration is taken into account not only for the data-collection period of the experiment, but in some instances, for materials handling and storage as well. In this section, we discuss the particular needs of \ndbd and natural source experiments. 



\subsection{Infrastructure}
Infrastructure needs of \ndbd and natural source experiments are based on long-term occupation of very clean underground environments. Many detector technologies also require stable long-term cryogenic operations, and most require remote computing, and safe and efficient site access for investigators. 

\paragraph{Space and Depth}

 \ndbd and low energy neutrino studies have to be conducted deep underground in clean environments.  Current experiments operate at depths on the order of 1 km or greater \cite{U1-henning2016}, including the CUORE experiment at 3400 m.w.e, and SNO+ at 6000 m.w.e.   In the United States, two tonne-scale experiments (LEGEND1000 \& nEXO) have both listed SNOLAB site (depths of up to 6000 m.w.e.) as the preferred location\textemdash - an indication of the high value placed on deep underground space by these experiments, and of the critical need for additional space to be developed for next-generation experiments.  Studies regarding the depth requirements for tonne and multi-tonne scale detectors are still underway.  
\paragraph{Power and Electrical} 
Underground electrical systems for live experiments and detector testing should provide clean power and fail-overs to backup power in case of a power outage. Cryogenic coolers (e.g. pulse tube cryocoolers) and dilution refrigerators rely on a continuous supply of power for operation.  The sudden interruption of power can very quickly lead to equipment damage and in the worst-case scenarios, rapid boil-off of liquid cryogens, loss of data, and physical safety hazards. 

The type of backup power used may vary based on availability and experimental demands.  Cryogenic experiments frequently use a combination of backup generator power and one or more uninterruptible power supplies (UPS).  A generator that kicks in within a few seconds of a power outage can supply enough power to maintain large equipment (such as a pulse tube cryocooler), while a sufficiently sized UPS will provide enough energy to keep computers and instrumentation running in the interval between primary power loss and switchover to the generator.  The rapidity with which the swichover can be accomplished depends on whether the generator operates continuously; for low background experiments, a rapid turnover is required to prevent shutdown of cooling equipment and computers that could lead to loss of data and equipment damage. 

In addition to backup power, sensitive electronic measurements (e.g. $\mathcal{O}(\si{\nano\volt})$ and $\mathcal{O}(\si{\pico\ampere})$ signals) can be compromised by electronic noise, including spurious noise on AC power lines. Clean power is used to obtain a clean sinusoidal signal for such measurements.  One method of providing clean power is through use of a UPS, which receives power from the general AC source and rectifies and corrects the output voltage provided to the electronic instrumentation. While dilution refrigerators and cryocoolers do not usually require clean power, the devices under test in those systems often do, and this should be considered when planning power for instrumentation.

To facilitate the design and construction of experiments, the underground location should be well-characterized. This includes 3D geographical scans of the cavern, rock sampling and characterization of the radioactive backgrounds of the area. 

Many \ndbd and natural source experiments will also require gas consumables, e.g., liquid nitrogen or N$_2$-boil off. These facilities could be shared in underground locations. 

\subsection{Cleanrooms}
All \ndbd and natural source experiments will require long-term occupation of dedicated cleanrooms, typically for the full period of installation and operation of the experiment. However, the cleanroom requirements vary per experiment. The cleanliness spec is from Class-1(ISO-3) to Class-10000(ISO-7), with a surface area of 50--200\,m$^2$ and 3--10\,m headroom. The radon requirements in the cleanroom also range from a very challenging 1\,mBq/m$^3$ to a more modest 10\,Bq/m$^3$. The lower limit is only achievable with dedicated Rn-abatement systems or synthetic air. Some of the cleanrooms will need to be equipped with gloveboxes to further reduce the Rn concentration and/or fume hoods for chemical procedures. Humidity and temperature control are essential. 

Given the long-term occupation and varying requirements, most \ndbd and natural source experiments will need their own dedicated cleanrooms. Shared cleanrooms could be useful for additional short-term needs, such as very high cleanliness specification during the assembly of critical components. 




\subsection{Muon veto systems}
The depth requirement for \ndbd and natural neutrino source experiments is based on the need to shield the detectors from cosmic rays. These particles, which are often highly relativistic muons and protons, deposit energy into the detector and trigger as events, raising the temperature and noise levels and increasing the proportion of ``dead time'' during which the detector is not sensitive to new events.  

The muon veto serves as a complement to the primary detector.  While the rock overburden of the underground detector site serves to attenuate the muon flux, the muon veto acts to actually tag muons that have reached the underground cavern and remove the associated events from analysis, improving the sensitivity of searches for rare events like \ndbd and dark matter by identifying and removing backgrounds from the region of interest. 

Current and upcoming experiments installing a muon veto, such as CUORE and CUPID \cite{U1-cupidprecdr}, have taken advantage of developments in data acquisition hardware, including compact form factors and increased channel counts, to maintain a minimal footprint relative to the primary detector, so that such systems add little to the required space for an experiment.  Simulations by other experiments indicate that muon vetos can achieve rejection efficiencies of ~99.9\%, and background reduction of up to two orders of magnitude in the region of interest \cite{U1-wiesinger2018}.


This is consistent with the background reduction achieved by experiments such as \textsc{Gerda}, which reported a 99.9\% muon rejection efficiency and a reduction in the background index in the region of interest of nearly two orders of magnitude \cite{U1-wiesinger2018}.


\subsection{Clean environments for materials}
Commissioning of large low-background experiments requires careful materials handling protocols to maintain the radiopurity of the materials and detectors during assembly. Production of clean materials directly underground mitigates some of the cost and logistical complications associated with transporting materials above ground and minimizes cosmogenic activation that might occur during transit.  
The copper used in shielding and cryogenic infrastructure often directly faces active detector volumes in low-background experiments, and is often the source of non-negligible backgrounds due to such activation \cite{U1-she2021, U1-baudis2015}.  
Underground electroformation facilities provide a means of producing copper with minimal surface exposure, resulting in lower backgrounds. 

Underground argon facilities are an example of the supporting resources needed for large low background experiments.  Underground facilities for the extraction, processing, and storage of argon to prevent cosmogenic activation through handling on earth's surface.

Similar considerations exist for the crystals used in bolometric detectors used several of the largest neutrino, \ndbd, and dark matter detectors.  As experiments move to tonne and multi-tonne scale detector masses, a main source of backgrounds arises from cosmogenic activation of the detector material \cite{U1-cosmoact2016}.
As a particular example, while Ge detectors lead the world in detection thresholds for dark matter searches and have excellent energy resolution in discriminating $2\nu\beta\beta$ events, providing discovery potential for \ndbd searches, the cosmogenic production of tritium,  $^{68}\textrm{Ge}$ and $^{60}\textrm{Co}$ are a main source of background for dark matter experiments and constrain the sensitivity of Ge \ndbd experiments beyond the scale of LEGEND \cite{U1-dm2022pc}, and may be a significant source of backgrounds in next-generation tonne-scale experiments in other isotopes (e.g. CUPID-1T) as well.  

A laboratory for underground crystal growth could be located at depths ranging from 300-ft at the SURF to very deep at SNOLAB, as long as the hadronic components of cosmic rays are significantly reduced. Such a laboratory should include zone refining for Ge ingots, and detector fabrication facilities. In addition, a mechanical lab should be attached to the crystal growth facility because the mechanical process of crystals is part of the entire production chain. Locating the entire production chain underground at the underground labs where the experiments will be built will provide significant reduction in the cosmogenic production of radioactive isotopes \cite{U1-dm2022pc}.
An underground facility for crystal growth and fabrication requires sufficient space (4 labs for a total of ~1000 square feet), underground safety in terms of exhausting hydrogen gas (ambient gas for processing Ge in zone refining and crystal growth) and handling the waste of acids (clean and etching of Ge). However, there are safety protocols that we have established and implemented for more than 10 years at the surface labs. Those challenges can be overcome \cite{U1-dm2022pc}. 

Production of deionized ultra-pure water underground is also needed not only for cleaning but in order to fill muon-veto water tanks, and means of production in sufficient quantities is a facilities issue that should be considered with other infrastructure needs and background mitigation strategies.

Space and facilities are also needed for handling of fluids such as organic liquid scintillator, including filling, recirculation, and purification capabilities.

\subsection{Clean environments for detector construction}
Of the respondents to the survey, the most concentrated use of cleanroom environments was for detector assembly and construction, with a few experiments also expressing continued need for cleanroom space for the purposes of calibration/operations, supplementary measurements, parts cleaning, and clean storage.  During the construction and assembly phases, clean space is used to prevent contamination of parts with radon, a long-lived radioactive daughter of the U/Th chain. Cleanroom classes particularly suitable for these needs are ISO5, ISO 6, and radon-free clean rooms.   Experiments that can be assembled from smaller modules typically reported plans to assemble detectors in glove boxes, mitigating the need for special requirements from the facility. The few responding experiments with specified environmental radon levels typically reported requirements on the order of \SI{1}{\milli\becquerel\per\meter\cubed}, with one experiment specifying a flow of 180-220 \si{meter\cubed\per\hour}.  KamLAND-Zen requires \SI{10}{\becquerel\per\meter\cubed} or better.  

\subsection{Material Assay Facilities}
In general, neutrino experiments constructed and installed in underground laboratories are highly sensitive to radioactive backgrounds.  Background budgets and projections rely on accurately quantifying the expected background rates, including though material assay of the detector components. 
A variety of assay techniques are employed by neutrino experiments. High-purity germanium (\hpge) detectors are commonly used by individual institutions and experiments, and as a result there is little underground-lab-supplied capacity for \hpge{} measurements. Existing facilities are usually oversubscribed. Dedicated assay facilities play a larger role in measurements involving alpha-counting, inductively-coupled plasma mass spectrometry (\icpms), 
neutron activation, and dedicated cryogenic bolometers for characterizations requiring sensitivity on the order of \si{\micro\becquerel\per\kilogram}. Future neutrino experiments will need to screen 10s to 100s of samples a year and compete for assay facilities with dark matter experiments requiring similar screening efforts.

\subsection{Storage facilities}
The 2013 report contained a recommendation to reserve underground space for materials assay and storage 
\cite{U1-css2013ulc}. This is a continued need for experiments with stringent constraints on radiopurity.

Underground facilities for storage continue to play a role in mitigating cosmogenic activation, including muon-activation, of materials.

\subsection{Environmental monitoring and safety}

The growing scale of underground neutrino experiments requires the development of clear strategies for environmental monitoring and safety protocols to maintain sustainable operations.  Sensitive radon detectors would provide ongoing information regarding levels relevant to experimental backgrounds and human health 
Several experiments 
rely on storing large amounts of liquid cryogens underground, which carries hazards associated with handling extremely cold materials and asphyxiation. Since the experiments are run in underground caverns, they are often located in active mines, and must be coordinated with mining activity.  Associated with this are challenges involving ease of access for equipment and personnel, and ease of egress or access to refuge chambers in the event of an emergency. 






\subsection{Underground Testing and R\&D Facilities}
Large-scale experiments\textemdash particularly those planned for next-generation very-low-background neutrino and \ndbd searches\textemdash often require long-term research and development and prototyping efforts to characterize sensors and detector assemblies.  For neutrino experiments, full testing of low background detectors requires that the R\&D setup needs to be shielded from cosmic rays and placed underground. The space requirements for such setups are typically modest in comparison to the actual experiment and do not necessarily have to be in the same underground laboratory.  Particularly in the context of international collaborations, this frees up restrictions on location, and is often preferred when proximity to the home institution is desired. 

Because of the integral role of quantum sensing in detector development for next-level fundamental physics searches, and the use of similar devices in quantum computing and information science, underground test facilities will also provide opportunities for collaboration and cooperation with QIS.  Synergistic arrangements, including shared facilities, could benefit both fields.  

Typical instrumentation for an underground test facility might include cryogenic support, including dilution refrigeration.

\subsection{Underground User Facilities and International Collaboration}
The previous P5 report placed strong emphasis on the need for international collaboration \cite{U1-mpsac-p5-2015}.
While the collaborative nature of neutrino physics requires cooperation from scientists around the world, many face practical barriers, including lack of computing account access for foreign-national collaborators from ``countries of concern''.
SURF is applying to be a DOE user facility \cite{U1-doenuf}.  There is strong community support for such facilities to improve access to on-site and computing resources to encourage international collaboration.

\subsection{Other considerations: Domestic Impacts and Broader Participation}
The collaborative nature of astro- and geophysically-sourced neutrino and \ndbd experiments is well suited to address the MPSAC 2015 recommendation \cite{U1-mpsac-p5-2015} to strengthen relationships between laboratories and universities to provide training and harness the unique capabilities of each.  Over the last decade, the investments made in graduate programs and laboratory research in instrumentation, hardware, and data science computing have yielded enormous benefits 
within and far beyond particle physics.   

\subsection{Goals}
Major goals of the neutrinoless double-beta decay program include establishing full coverage and beyond of the Inverted Ordering mass region.  
Several major goals are incorporated within the scope of the underground frontier, and overlap with the needs of the cosmic frontier and the search for new physics. 
Neutrinos from natural sources are used both to probe the source, and to study neutrino properties.  In both cases, the goals for this frontier are to establish and provide sufficient underground space and facilities to support these programs into the next decade and beyond.



\subsection{Synergies with Dark Matter Experiments and Quantum Information Science}
\subsubsection{Synergies with Dark Matter}
Neutrino experiments share many of the same radioactivity and activation requirements as direct detection dark matter (DM) experiments. Here also lie several synergies between neutrino experiments themselves, but also with the DM community. Underground facilities could take a lead on further fostering these interactions. 

There are already efforts underway to collect and store material assay results in a publicly accessible database, e.g., radiopurity.org. These efforts could be further strengthened so that experiments can more easily decide on construction materials. However, the usefulness of the stored information is often complicated due to, e.g., large batch to batch screening variations for the same material (even from the same manufacturer) or possibly for (scientific) competitiveness reasons. In addition, standardization of some screening methodologies, such as for Rn and plate-out measurements is required in order to compare across different labs and facilities. 

Activation studies is another area where neutrino and DM experiments could share expertise. These require the collection of various muon and neutron-related cross sections and tools to analyze them. There are virtually no community tools and databases at the moment and also here it seems that every experimental collaboration starts anew. 

Another area of synergy is in the use of underground test facilities for cryogenic and liquid scintillation detectors (see also Section \ref{futureDMexperiments}).  Research and development on these detectors is often impossible in surface laboratories due to high background rates. 

\subsubsection{Synergies with Quantum Information Science}

The overlap of the facilities needs of \ndbd and natural source experiments with those of quantum information science is largely in the area of sensor development and in the shared use of underground space and infrastructure for cryogenic tests of prototypes and detector R\&D.  The synergistic aspects of these fields are more extensively discussed in Section \ref{sec:qisynergy}.




\section{Conclusions}
Conclusions from the report are summarized below.  
\begin{itemize}
\item Preservation of competitiveness in the US neutrino program is dependent on future access to clean, deep underground space.  Planning for future needs must begin now.

\item Local, expert support at underground facilities is critical for success.

Guidance and local expertise on site-specific deployment details, environmental health and safety requirements, fluid handling, shipping, and administration are vital components of executing and maintaining operations in underground experiments (especially those of larger scales)

\item Supporting underground facilities is critical to achieving science goals for neutrino physics experiments

Several common themes arise with respect to the resources and facilities needed for successful underground neutrino physics.  This includes capabilities that are not unique to these experiments but are critical in assessing underground-specific needs such as depth, environmental backgrounds, materials handling and storage (and whether items must be prepared and stored underground - space needs) and cleanliness requirements:
\begin{itemize}
    \item There is a need for central coordination of radio assay capabilities, perhaps by one of the underground laboratories.
    \item There is strong community support for better user facility support for international collaborators.
    \item Robust computing/connectivity in labs underground;
    \item Centralised data for cosmogenic activation;
    \item Centralised radiopurity database;
    \item Shared data and simulations;
    \item Facilities for radio-assays, low-background counting;
    \item Support for laboratory access for equipment and personnel;
    \item Support to understand and implement seismic safety requirements;
    \item Additional supporting facilities such as space for prototypes, R\&D etc.  A need for cryogenic facilities has synergies with dark matter and QIS.
\end{itemize}

\item Plans to pursue neutrino science beyond the tonne-scale \ndbd program require inter-agency planning efforts for future underground facilities, particularly with respect to the preparation and allocation of deep underground space.

\item There is a need to compile results from \ndbd, G2 dark matter, and natural source experiments, and to perform simulations regarding the sufficiency of depth of existing laboratories to host future-generation experiments.

As referenced in the 2013 report \cite{U1-css2013ulc}, depth requirements for neutrino and dark matter experiments depends on which technology is employed.  In keeping with the several avenues of research and development currently being pursued by \ndbd experiments, simulations corresponding to suitable combinations in each major category (e.g. detector substrate, sensor, etc) may also inform whether current depth and experimental space  constraints require more concentrated development of a particular technology for a planned experiment to be sited, or if underground facilities must be expanded.


\end{itemize}




